\documentclass[fleqn,usenatbib]{mnras}
 \pdfoutput=1 
\usepackage{newtxtext,newtxmath}

\usepackage[T1]{fontenc}

\DeclareRobustCommand{\VAN}[3]{#2}
\let\VANthebibliography\thebibliography
\def\thebibliography{\DeclareRobustCommand{\VAN}[3]{##3}\VANthebibliography}

\usepackage{float}
\restylefloat{table}
\restylefloat{figure}
\usepackage{graphicx}	
\usepackage{caption}
\usepackage{amsmath}	

\title[Learned Emulation of Radiative Transfer]{Enhancing 3D Planetary Atmosphere Simulations with a Surrogate Radiative Transfer Model}

\author[Tara P. A. Tahseen]{
Tara P. A. Tahseen,$^{1}$\thanks{E-mail: tara.tahseen.22@ucl.ac.uk (TPAT)}
João M. Mendonça,$^{2,3,4}$
Kai Hou Yip$^{1}$
and Ingo P. Waldmann$^{1}$
\\
$^{1}$Department of Physics and Astronomy, University College London, Gower Street, WC1E 6BT London, United Kingdom\\
$^{2}$Department of Space Research and Space Technology, Technical University of Denmark, Elektrovej 328, 2800 Kgs.\ Lyngby, Denmark\\
$^{3}$ Department of Physics and Astronomy, University of Southampton, Highfield, Southampton SO17 1BJ, UK \\
$^{4}$ School of Ocean and Earth Science, University of Southampton, Southampton, SO14 3ZH, UK}

\date{Accepted 2024 October 27. Received 2024 October 18; in original form 2024 July 03}

\pubyear{\the\year{}}

\begin{document}
\label{firstpage}
\pagerange{\pageref{firstpage}--\pageref{lastpage}}
\maketitle

\begin{abstract}
This work introduces an approach to enhancing the computational efficiency of 3D atmospheric simulations by integrating a machine-learned surrogate model into the OASIS global circulation model (GCM). Traditional GCMs, which are based on repeatedly numerically integrating physical equations governing atmospheric processes across a series of time-steps, are time-intensive, leading to compromises in spatial and temporal resolution of simulations. This research improves upon this limitation, enabling higher resolution simulations within practical timeframes. Speeding up 3D simulations holds significant implications in multiple domains. Firstly, it facilitates the integration of 3D models into exoplanet inference pipelines, allowing for robust characterisation of exoplanets from a previously unseen wealth of data anticipated from JWST and post-JWST instruments. Secondly, acceleration of 3D models will enable higher resolution atmospheric simulations of Earth and Solar System planets, enabling more detailed insights into their atmospheric physics and chemistry. Our method replaces the radiative transfer module in OASIS with a recurrent neural network-based model trained on simulation inputs and outputs. Radiative transfer is typically one of the slowest components of a GCM, thus providing the largest scope for overall model speed-up. The surrogate model was trained and tested on the specific test case of the Venusian atmosphere, to benchmark the utility of this approach in the case of non-terrestrial atmospheres. This approach yields promising results, with the surrogate-integrated GCM demonstrating above 99.0\% accuracy and 147 factor GPU speed-up of the entire simulation compared to using the matched original GCM under Venus-like conditions.
\end{abstract}




\section{Introduction}

3D atmospheric models, commonly known as \textit{global circulation models} (GCMs) are key tools for studying the climates of solar system planets including the Earth, and are becoming increasingly important in the characterisation of exoplanet atmospheres. GCMs consist of multiple components which individually model different atmospheric processes: each component numerically solves equations governing an atmospheric process across elements of a grid and across many time-steps until simulation convergence criteria are reached. Due to the large number of numerical integrations involved in GCM simulations, producing a simulated atmospheric state for a given input set of planetary and stellar parameters is incredibly computationally expensive and time-consuming.


GCMs are applied in exoplanet science in the forward modelling component of Bayesian atmospheric retrieval pipelines. To retrieve posterior distributions on planetary and stellar parameters from a single transit spectrum, the Bayesian retrieval framework requires of the order of tens of thousands of forward simulations corresponding to different samples of input parameter space. With current state-of-the-art 3D modelling techniques, the compute resources and time needed to produce the required number of 3D simulations prohibits statistically-rigorous inference of data from the James Webb Space Telescope \citep[JWST,][]{gardner_james_2006} and further next-generation observational instruments yet to come, namely the Ariel Space Telescope \citep{tinetti_ariel_2021}. Acceleration without compromising the accuracy of GCMs is thus one clear method of facilitating inference using JWST and post-JWST data in exoplanet science.

Reducing GCM simulation time would also incur benefits in climate science of Earth and other solar system planets. Increased speed of computation would enable simulations to be run at greater resolution, and/or with more physics included, thus enabling more realistic climate simulations to be achieved.

The past few years have yielded work in both exoplanet science and Earth climate science to accelerate 3D models. Much of this work in exoplanet science involves extending 1D models with extra parameters to reflect certain 3D atmospheric variations deemed necessary to account for 3D effects in phase-curve data \citep{changeat_taurex3_2020, irwin_25-d_2020, chubb_exoplanet_2022, himes_towards_2023, nixon_aura-3d_2022, feng_2d_2020}. This approach is prone to introducing biases to the simulated atmospheric states produced, and a fully 3D model parametrisation is essential to ensuring that simulations are robust against both known and unknown biases resulting from model oversimplifications \cite{pluriel_strong_2020}. Earth climate science, benefiting from a wealth of high-resolution observational measurements, has had a different set of methods employed, namely machine-learned surrogate models trained on such observations \citep{yao_physics-incorporated_2023, ukkonen_exploring_2022}. Surrogate models have not been broadly explored outside of Earth climate science, but demonstrate the potential of speed-up without requiring an oversimplified model parametrisation \citep{yao_physics-incorporated_2023, ukkonen_exploring_2022}. 

Of the components within GCMs, radiative transfer is often the slowest or least-resolved process. In the OASIS GCM \citep{mendonca_new_2014}, the radiative transfer component contributes to 50-99\% of the total simulation runtime (depending on the temporal resolution of the radiative transfer, see section \ref{sec:simulation-runtime} for more details) for massive and complex atmospheres such as Venus \citep{mendonca_modelling_2020}. There is thus key scope to substantially improve GCM computational efficiency by targeting the radiative transfer component specifically. 

To assess the effectiveness of our new approach, we are testing it on Venus' atmospheric conditions. Simulating the Venus atmosphere in 3D is very computationally intensive due to the need for a complex radiative transfer model to accurately represent the energy balance in the atmosphere  \citep[e.g.][]{eymet_net_2009, lee_angular_2012, mendonca_new_2015}. Venus has a substantial amount of CO$_2$, which generates a strong greenhouse effect \citep{sagan_physical_1962} and is covered by highly reflective sulfuric acid clouds, obscuring the planet's surface. Only about 2.5$\%$ of the incoming solar radiation reaches the surface  \citep[e.g.][]{tomasko_measurements_1980, mendonca_new_2015}. The radiation model also requires fine spectral resolution to capture the spectral windows impacting energy exchange between the deep atmosphere and the upper layers above the clouds. Additionally, due to the high thermal inertia of the massive CO$_2$ atmosphere, the models need to be integrated over a long period to reach a statistically steady state \citep{mendonca_exploring_2016}. Using a surrogate model to represent the radiative transfer is key to enhancing the performance of 3D simulations whilst enabling a more realistic depiction of radiative processes at minimal cost. Therefore, Venus presents a significant challenge and serves as a benchmark for the complexities our new modelling approach may encounter in future applications.
\section{Data \& Codes}

\subsection{OASIS}\label{oasis}

OASIS is a planetary climate model composed of different coupled modules representing physical and chemical processes within planetary atmospheres \citep{mendonca_modelling_2020}. The mathematics and assumptions of the radiative transfer component of OASIS are detailed in \cite{mendonca_new_2015}. 

The equations in OASIS are discretized over concentric icosahedral spatial grids \citep{mendonca_thor_2016,deitrick_thor_2020}. For the 3D simulations in this study, we set the grid to approximately 2 degrees in the horizontal and 49 vertical layers. This grid configuration results in a total of 10242 columns covering the entire model domain. Our simulations started from the converged state obtained in \cite{mendonca_modelling_2020} and were integrated for 5 Venus solar days (approximately 117 Earth days) using a time step of 15 seconds. The model configuration is similar to the simulations in \cite{mendonca_modelling_2020}, and in the following sections, we describe the main physical modules relevant to this work.

\subsubsection{OASIS-RT}\label{oasis-rt}


The radiative transfer module of OASIS (henceforth referred to as OASIS-RT) models the interaction of radiation with gas and cloud species in the atmosphere in a two-stream manner. Radiation from two different sources are modelled separately: these are \textit{solar} radiation ($0.1-5.5 \ \mu \text{m}$) and \textit{thermal} radiation ($1.7 - 260 \ \mu \text{m}$). The solar radiation code utilises the $\delta$-Eddington approximation \citep{joseph_delta-eddington_1976} in combination with an adding-layer method \citep{liu_advanced_2006, mendonca_new_2015}, whilst the thermal code considers absorptivity and emissivity \citep{mendonca_new_2015}. The radiation scheme uses the $k$-distribution method to represent the gas absorption cross sections, which are integrated over 353 spectral bands and 20 Gaussian points \citep{mendonca_modelling_2020}. 

In the OASIS code used for this project, the spatial distribution and radiative properties of the clouds, which are composed of sulphuric acid and water, were taken from \cite{crisp_radiative_1986} and \cite{mendonca_new_2015}. The main cloud deck is located between roughly $45-65 \, \text{km}$ altitude with layers of sub-micron particles below and above \citep{knollenberg_microphysics_1980}. More details on the cloud properties can be found in \cite{mendonca_new_2015}.

The radiative transfer code involves two steps of computation for each timestep; these are:
\begin{enumerate}
    \item \textbf{Computing the gas optics:} computing the optical properties of layer boundaries from the thermodynamic variables. 
    \item \textbf{Computing the flow of radiation}: computing the upward and downward-welling fluxes at each layer boundary, from the incident flux at the top of the column in combination with the optical properties at each layer boundary.
\end{enumerate}

The model uses a $k$-distribution table \citep{lacis_description_1991} to calculate the optical properties of each layer across the wavelength bands using pre-computed wavelength-dependent absorption coefficients for $\text{CO}_2$, $\text{SO}_2$ and $\text{H}_2\text{O}$. These coefficients are combined with a continuum absorption, mostly from $\text{CO}_2$–$\text{CO}_2$ collisions, and Rayleigh scattering from $\text{CO}_2$ and $\text{N}_2$ \citep{mendonca_new_2015}.

The radiative transfer model takes inputs of the density $\rho$, pressure $p$, temperature $T$ and chemical composition per grid element as outputted from the dynamical core \texttt{THOR}\footnote{The model component that governs the resolved 3D fluid flow evolution.} \citep{mendonca_thor_2016}. Heating rates are then calculated per grid element from fluxes outputted from the radiative transfer model, and these heating rates update the temperature profile of the atmosphere, which serves as input back into the dynamical core. Flux is used to compute the heating rate per layer according to equation \ref{eqn:heating-rate}.

\begin{equation}\label{eqn:heating-rate}
    \frac{dT}{dt} = \frac{1}{\rho c_p} \frac{dF^{net}}{dz} 
\end{equation}
where $dF^{net}$ is the spectral-integrated net radiative flux ($\text{W}\,\text{m}^{-2}$), $\rho$ is the atmospheric density ($\text{kg} \,\text{m}^{-3}$), and $c_p$ is the specific heat capacity at constant pressure (900 $\text{J}\,\text{kg}^{-1}\,\text{K}^{-1}$).
%
%

In order to improve the computational efficiency of 3D simulations with radiative transfer, Venus GCMs traditionally do not update the radiative fluxes from the solar and thermal schemes at every time step \citep{lebonnois_wave_2016, mendonca_exploring_2016, mendonca_modelling_2020}. In the 3D simulations with explicit radiative transfer used in this study, the fluxes calculated from the solar radiation scheme were updated every 2880 steps, and the thermal radiation fluxes were updated every 320 steps. These values are adjusted by the model user and are specific to Venus's simulations. Larger values could potentially cause model instabilities. Although this approach introduces some inaccuracies in the heating/cooling rates, the robustness of the simulation is not compromised because the composition of the atmosphere and clouds remains constant over time. With the efficiency of the new surrogate model described in the next section, we are now able to update radiative fluxes at every time step. To enhance the stability of the model with the new surrogate model, we also apply the three-step Adams-Bashforth method to the heating rate calculated from the surrogate radiative fluxes.

\subsection{Data}\label{data}
The data used for this project is 3D data simulated by OASIS corresponding to input parameters of Venus \citep{mendonca_modelling_2020}. The data used in this work is from a simulation of Venus using OASIS-RT. The data consists of 1,000 timesteps of simulation, with the time interval between consecutive timesteps within the recorded dataset, $\Delta t$, set to 12 hours. These data correspond to timesteps of a full simulation for which the radiative transfer update has been executed. The data covers the entire icosahedral grid with dimensions 10,242 columns $\times$ 49 layers.

At the sample level (per atmospheric column), the data comprises the following quantities: pressure $p$, temperature $T$, and gas density $\rho$, all defined per layer; upwelling short-wave flux $\text{F}^{\text{SW}, \uparrow}$, downwelling short-wave flux $\text{F}^{\text{SW}, \downarrow}$, upwelling long-wave flux $\text{F}^{\text{LW}, \uparrow}$, and downwelling long-wave flux $\text{F}^{\text{LW}, \downarrow}$, all defined per layer boundary; and cosine of the solar zenith angle $\mu$, short-wave surface albedo $\alpha_{\text{SW}}$, long-wave surface albedo $\alpha_{\text{SW}}$, and surface temperature $T_0$, all defined per column.

Figures illustrating the distribution of samples within the dataset as a function of $\cos\mu$ and surface temperature are contained within appendix section \ref{distribution-of-samples}.

\section{Methods}

\subsection{Surrogate Modelling}\label{surrogate-modelling}

Surrogate models, or \textit{emulators}, are approximate mathematical models which model outcomes of interest, whereby the emulator mechanism does not necessarily reflect the physical mechanism which produces the outcomes. Surrogate models are useful in cases where the physical mechanism producing such outcomes is not well-understood, or in cases where modelling the physical mechanism is excessively computational demanding; in the case of the latter, the aim is to produce a surrogate model which is computationally efficient compared to the physical model whilst maintaining accuracy. Machine learning provides a framework by which surrogate models can be produced: deep neural networks adhere to the Universal Approximation Theorem, and can (in theory) model any arbitrarily complex non-linear relationship, thus providing a function space whereby a suitable surrogate function is almost certain to exist \citep{ian_goodfellow_and_yoshua_bengio_and_aaron_courville_deep_2016}.

Research into surrogate modelling of exoplanetary atmospheres is relatively nascent \citep{himes_accurate_2022,  himes_towards_2023, unlu_reproducing_2023}. The use of surrogate models within Earth Climate Science, though still new, is much more established and well-explored \citep{yao_physics-incorporated_2023, mukkavilli_ai_2023, ukkonen_exploring_2022}. The development of surrogate models for atmospheric models of exoplanets can thus be informed by the development of such surrogate models for the parameter space of Earth.

An example where surrogate modelling has already proved valuable within 3D modelling of exoplanetary atmospheres is work by \cite{schneider_harnessing_2024}, who utilised DeepSets to achieve fast and accurate mixing of correlated-$k$ opacities. Their work exemplifies how machine learning can be leveraged to successfully speed up a single process of a GCM whilst retaining accuracy, thus establishing a basis for further exploration into the integration of surrogates within GCM frameworks. 

In this work, two surrogate models were produced for 3D modelling of Venus using OASIS: one surrogate model to emulate the short-wave radiative transfer schema, and one to emulate the long-wave radiative transfer schema (see \ref{oasis-rt} for details on the computations for the two radiation schemas). 

\subsection{Data Preprocessing}\label{data-preprocessing}

Data were preprocessed separately for the long-wave and short-wave regimes. Each model took input of two types of data: variables defined at the grid-element level, referred to as \textit{vector} variables of dimension $(n_{\text{columns}},) = (49,)$, and variables defined at the column-level, referred to as \textit{scalar} variables.

The short-wave surrogate model took scalar inputs of: surface temperature, $T_0$; gas density of the lowest-altitude layer, $\rho_S$; pressure of the lowest-altitude layer, $p_S$; cosine of the solar zenith angle $\mu$; and short-wave surface albedo $\alpha_{SW}$. The long-wave surrogate model took scalar inputs of: surface temperature, $T_0$; gas density of the lowest-altitude layer, $\rho_S$; pressure of the lowest-altitude layer, $p_S$; and long-wave surface albedo $\alpha_{LW}$.

Both models took input of the same vector variables, which were as follows: temperature, $T$; pressure $p$; and gas density $\rho$. Vector variables were scaled as:
\begin{equation}
x_{i,j} = \frac{\log_e(x_{i,j})}{\log_e(x_{i,0})}
\end{equation}
for $x\in \{T, p\}$, and 
\begin{equation}
x_{i,j} = \left(\frac{x_{i,j}}{x_{i,0}}\right)^{0.25}
\end{equation}
for $x=\rho$, for the $i$th column and $j$th atmospheric level.

Scalar variables were then re-scaled as 
\begin{equation}
x_{i}^{\text{scaled}} = \frac{x_{i} - \min\limits_{i\in S_{\text{train}}}{x_i}}{\max\limits_{i\in S_{\text{train}}}{x_i} - \min\limits_{i\in S_{\text{train}}}{x_i}}
\end{equation}
for $x \in \{T_0, p_0, \rho_0\}$ where and $S_{\text{train}}$ is the training set.

\begin{figure*}
\includegraphics[width=1\linewidth]{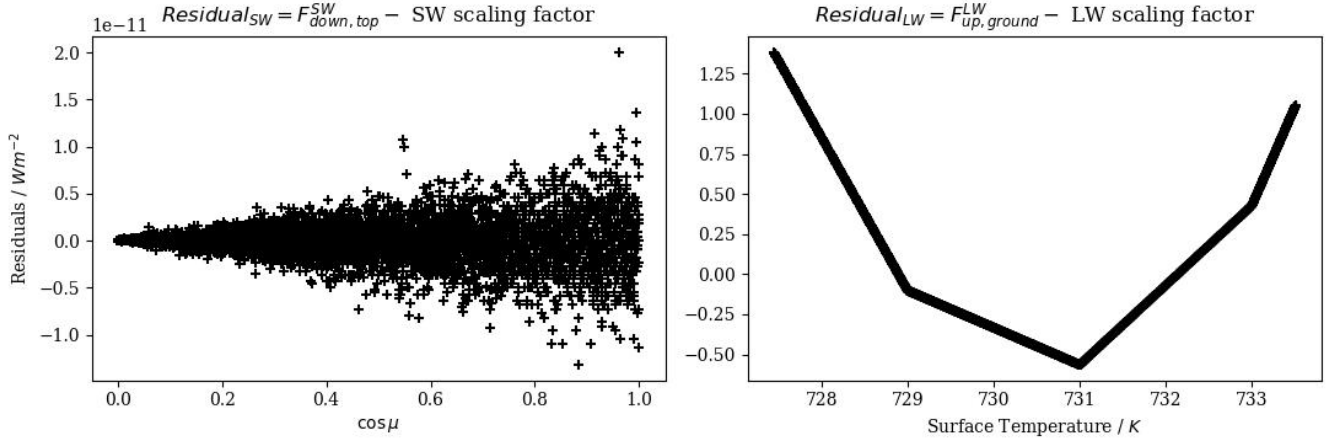}

\caption{The figures above display scatter plots illustrating the residuals between scaling factors \( A^{\text{LW}}(T_0) \) and \( B^{\text{SW}}(\mu) \) and their respective targets \( y^{\text{LW}, \uparrow}_{0} \) and \( y^{\text{SW}, \downarrow}_{49} \), plotted over one epoch of 10,242 test columns covering the entire icosahedral grid.
\textbf{Left:} This panel shows the residuals between the scaling factors \( B^{\text{SW}}(\mu_i) \) used for the short-wave targets of a given column \( i \) and the downward-welling short-wave flux at the top of the column \( \text{F}^{\text{SW}, \downarrow}_{i,49} \). \( B^{\text{SW}}(\mu_i) \), a linear function of the cosine of the solar zenith angle \( \mu_i \), approximates \( \text{F}^{\text{SW}, \downarrow}_{i,49} \).
\textbf{Right:} This panel displays the residuals between the scaling factors \( A^{\text{LW}}(T_{i,0}) \) used for the long-wave targets of a given column \( i \) and the upward-welling long-wave flux at the ground level \( \text{F}^{\text{LW}, \uparrow}_{i,0} \). Here, \( A^{\text{LW}}(T_{i,0}) \) approximates \( \text{F}^{\text{LW}, \uparrow}_{i,0} \) as a linear function of surface temperature \( T_{i,0} \).
\textbf{Both:} In both the long-wave and short-wave cases, the preferred quantities \( \text{F}^{\text{LW}, \uparrow}_{i,0} \) and \( \text{F}^{\text{SW}, \downarrow}_{i,49} \) to use for scaling flux profiles across atmospheric levels, involve complex calculations. The figures demonstrate that simple linear functions \( A^{\text{LW}}(T_0) \) and \( B^{\text{SW}}(\mu) \) yield close approximations with low residuals, making them suitable scaling factors instead.}
\label{fig:preprocessing-residuals}
\end{figure*}

Targets were scaled as follows: 
\begin{equation}
u^{\text{LW}}_{j} = \frac{y^{\text{LW}}_j}{A^{\text{LW}}}
\end{equation}
for $y_j \in \{ F_{j}^{\text{LW}, \uparrow}, F_{j}^{\text{LW}, \downarrow} \}$ and

\begin{equation}
u^{\text{SW}}_{j} = \frac{y_j^{\text{SW}}}{B^{\text{SW}}}
\end{equation}
for $y_j \in \{ F_{j}^{\text{SW}, \uparrow}, F_{j}^{\text{SW}, \downarrow} \}$, for the $j$th altitude level ($j \in [0,49]$ where 0 indexes the ground level and 49 indexes the top level of the atmospheric column), where $A^{\text{LW}}$ and $B^{\text{SW}}$ are scaling factors linear in $T_0$ and $\mu$ respectively, and fitted from the data:

\begin{equation}
A^{\text{LW}}(T_0) = a_1 \cdot T_0 + a_2
\label{eqn:sw-scaling-factor}
\end{equation}
\begin{equation}
B^{\text{SW}}(\mu) = b_1 \cdot \mu + b_2
\label{eqn:lw-scaling-factor}
\end{equation}

\noindent where $(a_1, a_2)$ are constants fitted using $y^{\text{LW}, \uparrow}_{0}$ and $(b_1, b_2)$ are fitted using $y^{\text{SW}, \downarrow}_{49}$ across all columns of the training set $S_{\text{train}}$. Values of $(a_1, a_2, b_1, b_2)$ are retained for model prediction post-processing. Residuals between the targets $y^{\text{LW}, \uparrow}_{0}$, $y^{\text{SW}, \downarrow}_{49}$, and the respective scaling factors approximating the value of these targets, are displayed in figure \ref{fig:preprocessing-residuals}. These simple linear scaling methods were chosen for data pre-processing in this work instead of using exact computations of $y^{\text{LW}, \uparrow}_{0}$ and $y^{\text{SW}, \downarrow}_{49}$ as the latter computations are much more involved, and the more simple computations produce results with an acceptably small marginal difference in the values of the fluxes.

Columns across all epochs were shuffled and split into train, test and validation datasets, in the ratio $70:15:15$. 

\subsection{Model Architecture}

\begin{figure*}
    \begin{centering}
    \includegraphics[width=1\linewidth]{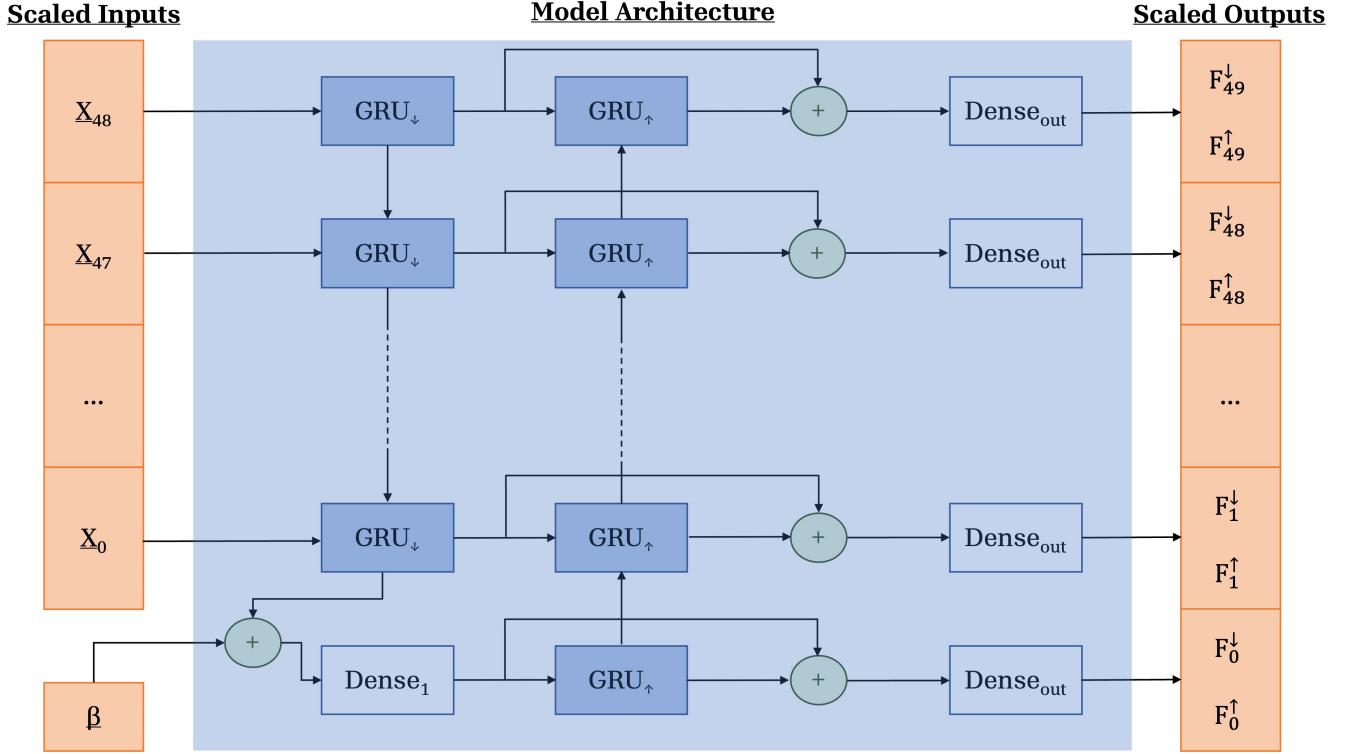}
    \end{centering}
    \caption{
    Figure illustrating the architecture of the learned surrogate models used in this work. Surrogate models for both the long-wave and short-wave schemas took scaled vector inputs $\vec{X}_j = (\tilde{p}_j, \tilde{T}_j, \tilde{\rho}_j)$ for the $j$th layer where $\tilde{p}_j, \tilde{T}_j, \tilde{\rho}_j$ are scaled pressure, scaled temperature and scaled density of the $j$th layer of the input atmospheric column, respectively (see \ref{data-preprocessing} for details on scaling). The surrogate models took input of scalar inputs $\vec{\beta}$, where $\vec{\beta}_{SW} = (\mu, p_0, T_0, \rho_0, \alpha_{SW})$ for the short-wave surrogate model, and $\vec{\beta}_{LW} = (p_0, T_0, \rho_0, \alpha_{LW})$ for the long-wave surrogate model, where $\mu$ and $\alpha$ denote the cosine of the solar zenith angle and surface albedo, respectively. Vector inputs are passed through a gated recurrent unit (GRU) layer starting from the top atmospheric layer (layer 48) and moving downwards to the ground layer (layer 0). Scalar inputs are concatenated with the output of the downwards GRU, and inputted into a dense layer, which then serves as the first input to an upward-moving GRU layer. The outputs of the two GRU layers are then concatenated and passed through a dense layer, which produces the model outputs of shape $(n_{\text{levels}}, n_{\text{outputs}}) = (50, 2)$, corresponding to scaled upwards and downwards flux for each atmospheric level. ReLU \citep{agarap_deep_2019} activation functions were used for both GRU layers and $\text{Dense}_1$, whilst a sigmoid activation function was used for the $\text{Dense}_{\text{out}}$ layer.
    }
    \label{fig:model-architecture}
\end{figure*}

Model architecture was chosen to be based on recurrent neural networks (RNNs)\footnote{See appendix \ref{appendix:model-architecture} for more details on neural networks and recurrent neural networks.}, as RNNs structurally incorporate the spatial dependence of the training data, which fits naturally in this scenario. RNN layers were implemented in the form of gated recurrent units \citep[GRUs; for further details on GRUs, see][]{cho_learning_2014}. Simple RNNs are susceptible to short-term memory problems, whereby information propagated forwards diminishes quickly. A common implementation of RNNs is the gated recurrent unit (GRU) which utilises a more sophisticated mechanism for propagating information “memory” forwards through a sequence: this is the base of the model architecture we employ in this work. The specific architecture of the benchmark was chosen to be that used by \cite{ukkonen_exploring_2022} (illustrated in figure \ref{fig:model-architecture}), which utilised a bi-directional RNN-based architecture to create a two-stream radiative transfer emulator for Earth, trained using observational data. Ukkonen's model performed with $\leq 0.5\%$ mean absolute error for the upwelling and downwelling fluxes on the test-set \citep{ukkonen_exploring_2022}, suggesting its potential efficacy for developing surrogates trained on analogous simulated data. The models used in this work were constructed and trained using TensorFlow version 2.12.0 \citep{abadi_tensorflow_2016}, and converted into ONNX format for integration within OASIS. Input data to the surrogate model is detailed in the table \ref{tab:inputs-table}, and surrogate model parameters are detailed in section \ref{appendix-model-parameters}.

\begin{table*}
\begin{center}
    \begin{tabular}{|c|c|c|} \hline 
         \textbf{Surrogate Schema}&  \textbf{Short-wave}& \textbf{Long-wave}\\ \hline 
         \textbf{Vector Inputs}&  $\vec{X}_i = (\tilde{p}_i, \tilde{T}_i, \tilde{\rho}_i)$& $\vec{X}_i = (\tilde{p}_i, \tilde{T}_i, \tilde{\rho}_i)$\\ \hline 
         \textbf{Scalar Inputs}&  $\vec{\beta}_{SW} = (\mu, p_0, T_0, \rho_0, \alpha_{SW})$& $\vec{\beta}_{LW} = (p_0, T_0, \rho_0, \alpha_{LW})$\\ \hline
    \end{tabular}
    \end{center}
    \caption{\textit{Surrogate model inputs:} Surrogate models for both the long-wave and short-wave schemas took scaled vector inputs $\vec{X}_i = (\tilde{p}_i, \tilde{T}_i, \tilde{\rho}_i)$ for the $i$th layer where $\tilde{p}_i, \tilde{T}_i, \tilde{\rho}_i$ are scaled pressure, scaled temperature and scaled density of the $i$th layer of the input atmospheric column, respectively (see \ref{data-preprocessing} for details on scaling). The surrogate models took input of scalar inputs $\vec{\beta}$, where $\vec{\beta}_{SW} = (\mu, p_0, T_0, \rho_0, \alpha_{SW})$ for the short-wave surrogate model, and $\vec{\beta}_{LW} = (p_0, T_0, \rho_0, \alpha_{LW})$ for the long-wave surrogate model, where $\mu$ and $\alpha$ represent the cosine of the solar zenith angle and surface albedo, respectively.}
    \label{tab:inputs-table}
\end{table*}
\subsection{Model Training}

Both models were trained in a supervised, end-to-end fashion. An Adam \citep{kingma_adam_2017} optimiser was used in combination with a cyclical learning rate. Models were trained using TensorFlow on an NVIDIA A100 GPU. 

\subsubsection{Loss Function}

The loss function was constructed as a combination of the mean percentage error between the predictions and targets, per output. The mean error (ME) for the $i$th test column and $k$th target variable is defined as follows:
\begin{equation}
    \text{ME}_{i, k} = \sum_{j=0}^{n_{\text{levels}}-1} \frac{\left| \hat{y}_{i,j,k} - \tilde{y}_{i,j,k}\right|}{n_{\text{levels}}}
    \label{MSE_ik}
\end{equation}
\noindent where $\hat{y}_{i,j,k}$ is the target for the $i$th test column, $j$th atmospheric level, and $k$th target variable, where $k=0$ corresponds to down-welling flux and $k=1$ corresponds to up-welling flux, and $n_{\text{levels}}$ is the number of atmospheric levels ($n_{\text{levels}}=50$ is this work). Normalisation factors were defined as 
\begin{equation}
    \text{norm}_{i,k} = \frac{\sum_{j=0}^{n_{\text{levels}}-1} \left| \hat{y}_{i,j,k} \right|}{n_{\text{levels}}}
\end{equation}
such that mean percentage error (MPE) of the $i$th test column and $k$th target variable can be expressed as 
\begin{equation}
\text{MPE}_{i,k} = \frac{\text{ME}_{i,k}}{\text{norm}_{i,k}}
\end{equation}
The loss function per sample was then defined as 
\begin{equation}
\text{loss}_i = \frac{1}{2} \sum_{k} \text{MPE}_{i,k}
\end{equation}
with the total loss defined as the sum over all test samples: 
\begin{equation}
\text{loss} = \frac{1}{2} \sum_i \sum_{k} \text{MPE}_{i,k}
\label{loss}
\end{equation}

\subsubsection{Hyperparameter Tuning}

Multiple models were trained corresponding to different hyperparameter values. Number of neurons of all RNN layers was varied across the range of values $[16, 32, 64, 128]$ for both surrogate models. The best candidate models were chosen as having 128 neurons per RNN layer for the short-wave surrogate model, and 32 neurons per RNN layer for the long-wave surrogate model.

\subsection{Performance Analysis}


Below, we detail the metrics used to analyse the performance of the surrogate models on the test set.
In the results (section \ref{results}), different aggregations of absolute error (equation \ref{scaled_AE_ijk} for raw model outputs and equation \ref{AE_ijk} for postprocessed model outputs) are used to investigate the performance of both the long-wave and short-wave surrogate models.

\begin{equation}
    \text{Absolute Error} \equiv \text{AE}_{i, j, k} = \left| \hat{y}_{i,j,k} - \tilde{y}_{i,j,k}\right|
    \label{AE_ijk}
\end{equation}
where $\hat{y}_{i,j,k}$ are the post-processed model predictions, and $\tilde{y}_{i,j,k}$ are the unscaled target variables.
\begin{equation}
    \widehat{\text{AE}}_{i, j, k} =  \left| \hat{u}_{i,j,k} - \tilde{u}_{i,j,k}\right|
    \label{scaled_AE_ijk}
\end{equation}
\noindent where $\hat{u}_{i,j,k}$ are the raw model predictions, and $\tilde{u}_{i,j,k}$ are target variables which have been scaled to lie in the interval $[0,1]$ using the scaling methods detailed in section \ref{data-preprocessing}.

Column-aggregated error quantities are defined as follows:

\begin{equation}
    \text{CAE}_{i, k} = \sum_{j=0}^{n_{\text{levels}}-1}\left| \hat{y}_{i,j,k} -\tilde{y}_{i,j,k}\right|
    \label{CAE_ik}
\end{equation}

\begin{equation}
    \widehat{\text{CAE}}_{i, k} = \sum_{j=0}^{n_{\text{levels}}-1}\left| \hat{u}_{i,j,k} - \tilde{u}_{i,j,k}\right|
    \label{scaled_CAE_ik}
\end{equation}
where $n_{\text{levels}}$ is the number of atmospheric levels.

Error quantities averaged across samples per altitude level are defined as follows:

\begin{equation}
    \text{MAE}_{j, k} = \frac{\sum_{i=0}^{N-1}\left| \hat{y}_{i,j,k} -\tilde{y}_{i,j,k}\right|}{N}
    \label{MAE_jk}
\end{equation}

\begin{equation}
    \widehat{\text{MAE}}_{j, k} = \frac{\sum_{i=0}^{N-1}\left| \hat{u}_{i,j,k} -\tilde{u}_{i,j,k}\right|}{N}
    \label{scaled_MAE_jk}
\end{equation}
where $N$ is the number of test samples.

Mean flux for the $k$th target variable is defined as:

\begin{equation}
\text{Mean Flux}_k \equiv \bar{F}_k = \frac{\sum_{i=0}^{N-1}\sum_{j=0}^{n_{\text{levels}}-1}{\tilde{y}_{i,j,k}}}{N}
\label{eqn:mean-flux}
\end{equation}
and the mean absolute error for the $k$th target variable aggregated across all altitude levels is calculated as 
\begin{equation}
    \text{MAE}_k = \sum_{j=0}^{n_{\text{levels}}-1}{\text{MAE}_{j,k}}
    \label{eqn:mae-k}
\end{equation}

\section{Results \& Discussion}\label{results}

\begin{table*}
\centering
\begin{tabular}{ |c|c|c|c|c| }
\hline
\textbf{Regime} & \textbf{Stream} & \textbf{Mean Flux $\bar{\textbf{F}}_k$ ($\text{W}\,\text{m}^{-2}$)}& $\textbf{MAE}_k$ ($\text{W}\,\text{m}^{-2}$)& $\textbf{MAE}_k$  / $\bar{\textbf{F}}_k$\\ 
\hline
Long-wave& Upwelling & 4211.0 & 18.8 & 0.45 \% \\ 
 & Downwelling & 4139.3 & 16.9 & 0.41 \% \\ 
 \hline
Short-wave& Upwelling & 577.9 & 6.4 & 1.11 \% \\ 
 & Downwelling & 707.8 & 7.7 & 1.09 \% \\ 
 \hline
\end{tabular}
\caption{The above table summarises the mean absolute error ($\text{MAE}_k$, defined in equation \ref{eqn:mae-k}) across the four target variables, and relative to the mean values $\bar{F}_k$ of these four variables (as defined in equation \ref{eqn:mean-flux}), across the test set.}

\label{error-table}
\end{table*}

\begin{figure*}
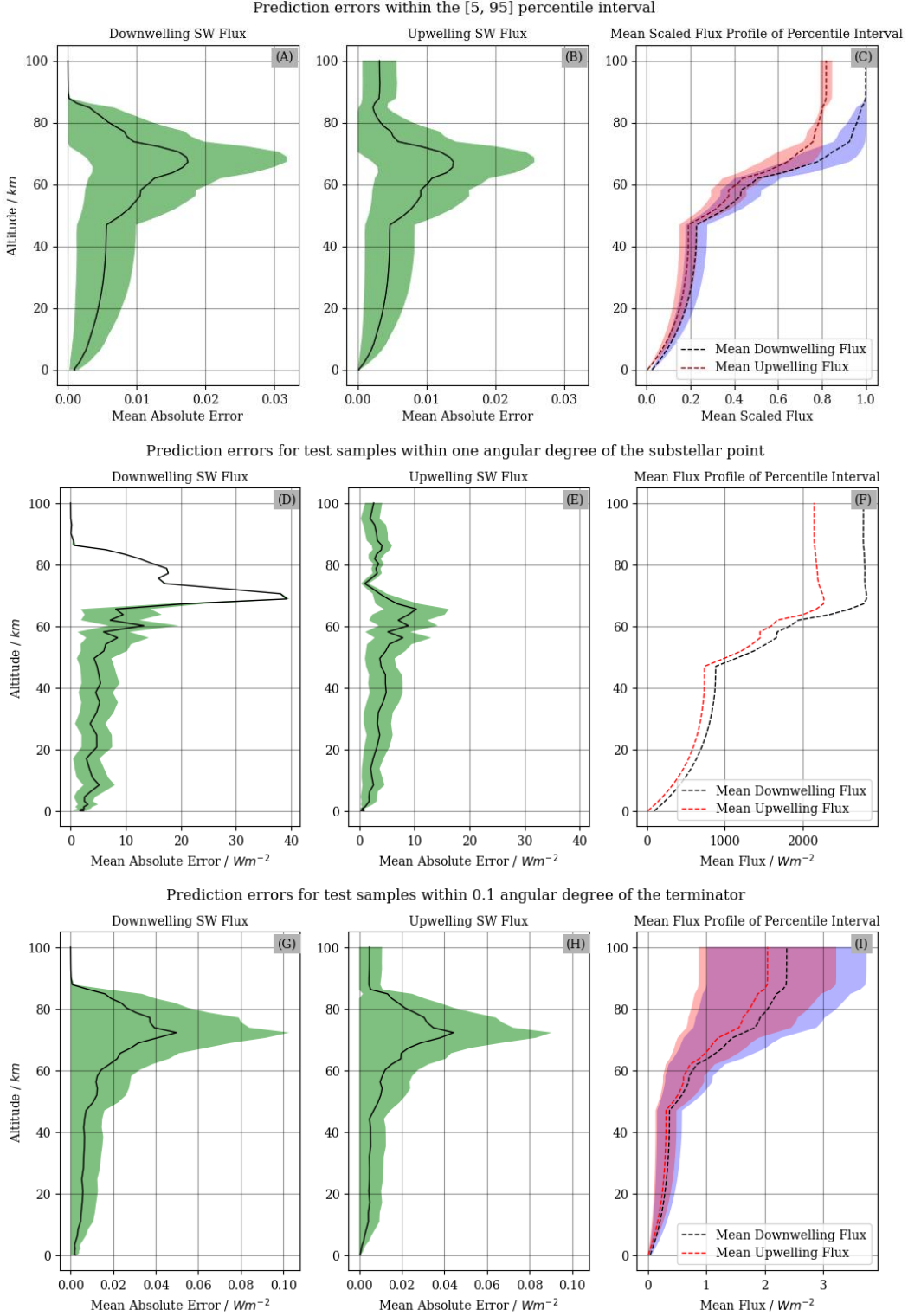

\centering
\hspace{3mm}\includegraphics[width=0.905\linewidth]{plots/new-paper-plots/03-june-2024/sw_scaled_mae_5_95.pdf}
\includegraphics[width=0.76\linewidth]{plots/new-paper-plots/01-jul-2024/err-substellar.pdf}\\
\includegraphics[width=0.76\linewidth]{plots/new-paper-plots/01-jul-2024/err-terminator.pdf}


\caption{Plots A and B display the $\widehat{\text{MAE}}_{j,k}$ (defined in equation \ref{scaled_MAE_jk}) for test samples which have column-aggregated absolute error ($\widehat{\text{CAE}}_{i,k}$, defined in equation \ref{scaled_CAE_ik}) in the $[5, 95]$ percentile interval (that is, test samples with errors falling in the smallest 5\% and largest 5\% of the test set have been discarded from this aggregate statistic, in order to better illustrate typical model performance). Plots D and E display the $\text{MAE}_{j,k}$ (defined in equation \ref{MAE_jk}) of surrogate model predictions across the two short-wave target variables, for test samples which lie within one angular degree of the substellar point; and plots G and H display the $\text{MAE}_{j,k}$ of surrogate model predictions across the two short-wave target variables, for test samples which lie within 0.1 angular degree of the day-night terminator. Plot C displays the scaled target short-wave flux profiles averaged across the test samples with $\widehat{\text{CAE}}_{i,k}$ within the chosen percentile interval. Plot F displays the target short-wave flux profiles averaged across samples which lie within one angular degree of the substellar point; and plot I displays displays the target short-wave flux profiles averaged across samples which lie within 1.1 angular degree of the terminator. The shaded regions in all plots represent the interval $[\max{(0,\theta_j-\sigma_j)},\,\, \theta_j+\sigma_j]$ where $\theta_j$ and $\sigma_j$ denote the mean and standard deviation of the plotted quantities for the $j$th altitude level.}
\label{fig:sw-mae}
\end{figure*}

\begin{figure*}
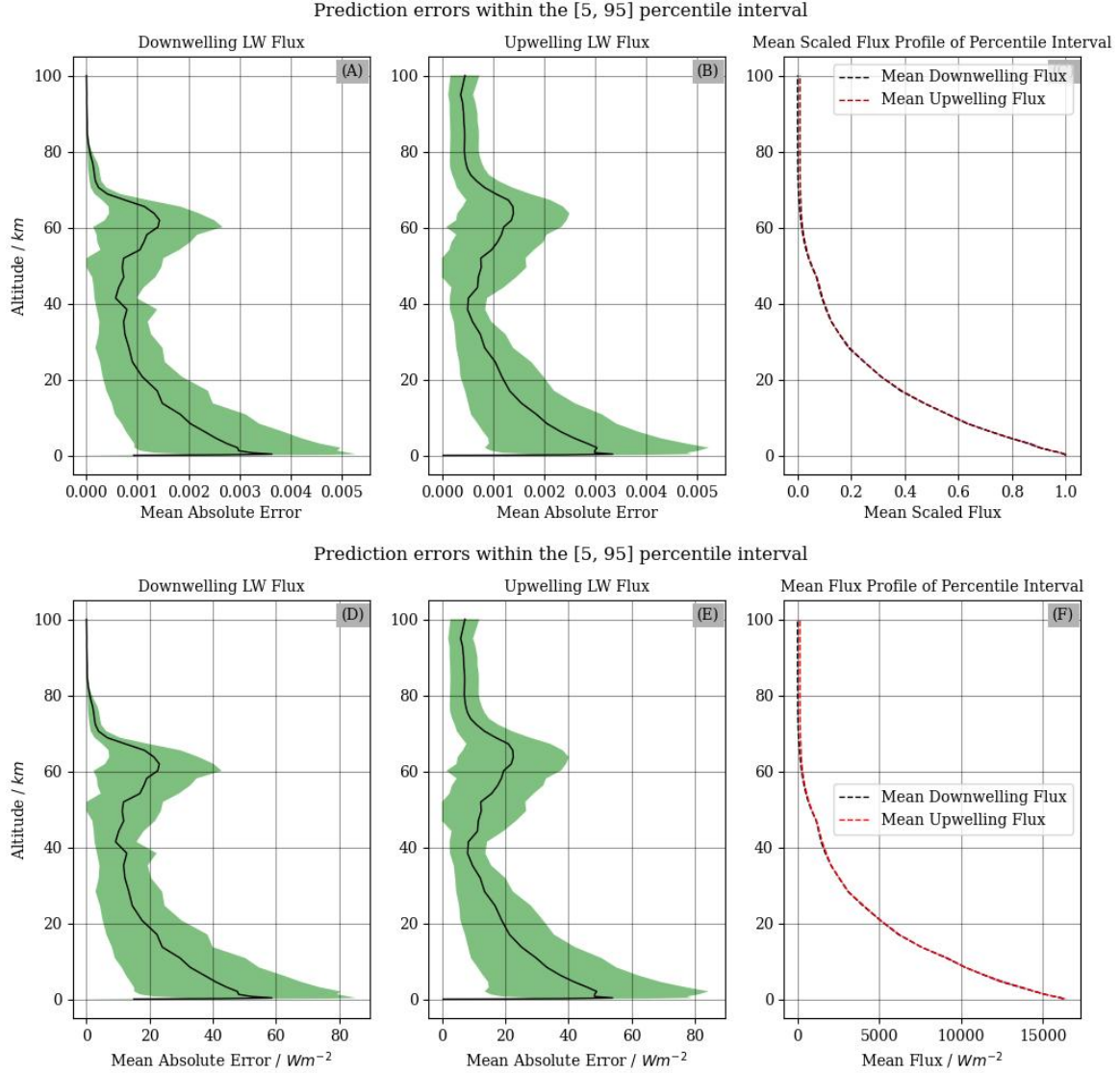

\includegraphics[width=1\linewidth]{plots/new-paper-plots/03-june-2024/lw_scaled_mae_5_95.pdf}
\includegraphics[width=1\linewidth]{plots/new-paper-plots/03-june-2024/lw_mae_5_95.pdf}
\caption{Plots A and B display the $\widehat{\text{MAE}}_{j,k}$ (defined in equation \ref{scaled_MAE_jk}) and plots D and E display the $\text{MAE}_{j,k}$ (defined in equation \ref{MAE_jk}) of surrogate model predictions across the two long-wave target variables, for test samples which have column-aggregated absolute error ($\widehat{\text{CAE}}_{i,k}$, $\text{CAE}_{i,k}$) in the $[5, 95]$ percentile interval (that is, test samples with errors falling in the smallest 5\% and largest 5\% of the test set have been discarded from this aggregate statistic, in order to better illustrate typical model performance). Plot C displays the scaled target long-wave flux profiles averaged across the test samples with $\widehat{\text{CAE}}_{i,k}$ within the chosen percentile interval. Plot F displays the target flux long-wave profiles averaged across the test samples with ${\text{CAE}}_{i,k}$ within the chosen percentile interval (not necessarily the same subset displayed in plot C). The shaded regions in all plots represent the interval $[\max{(0,\theta_j-\sigma_j)},\,\, \theta_j+\sigma_j]$ where $\theta_j$ and $\sigma_j$ denote the mean and standard deviation of the plotted quantities for the $j$th altitude level.}
\label{fig:lw-mae}
\end{figure*}

\begin{figure*}
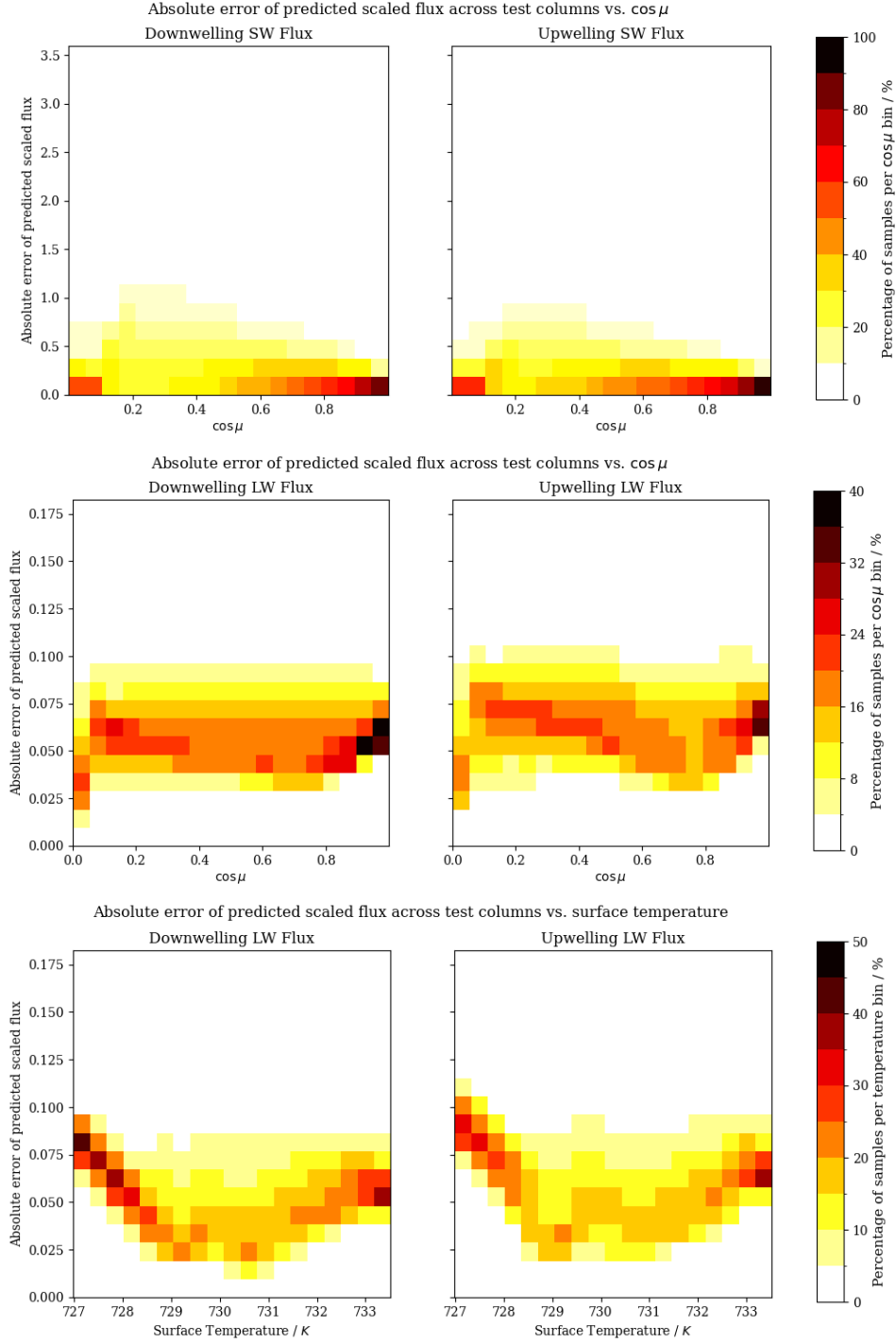

\centering
\hspace*{0.3cm}\includegraphics[width=0.7\textwidth]{plots/new-paper-plots/27-may-2024/sw-error-vs-cosz.pdf}
\includegraphics[width=0.7\textwidth]{plots/new-paper-plots/27-may-2024/lw-error-vs-cosz.pdf}
\includegraphics[width=0.7\textwidth]{plots/new-paper-plots/27-may-2024/lw-error-vs-stemperature.pdf}
\caption{The above plots display the variation in $\widehat{\text{CAE}}_{i,k}$ (as defined in equation \ref{scaled_CAE_ik}) across the long-wave and short-wave schemas, as functions of the cosine of the solar zenith angle ($\cos\mu$) and of surface temperature $T_0$ (for the long-wave schema only). The extent of the plots on the y-axis covers the full range of the error quantity on the y-axis of each plot, with white spaces corresponding to values falling in the lowest value bin.
\\\textbf{Top row:} The above figure displays the error distribution of test samples per $\cos\mu$ bin, across both short-wave target variables. Test samples have been binned into 20 bins of $\cos\mu$, into 20 bins of $\widehat{\text{CAE}}_{i,k}$. Bins have been normalised to percentages by the total number of test columns in each $\cos\mu$ bin.
\\\textbf{Middle \& bottom rows:} The above figure displays the error distribution of test samples per $\cos\mu$ bin, across both long-wave target variables as a function of $\cos\mu$ (middle row) and $T_0$ (bottom row). $\cos\mu$, $T_0$, and $\widehat{\text{CAE}}_{i,k}$ have been divided linearly into 20 bins; the plots display the proportion of test samples falling into each error bin relative to total samples in a given $\cos\mu$ or $T_0$ bin.}
\label{fig:error-variations}
\end{figure*}

\begin{figure*}
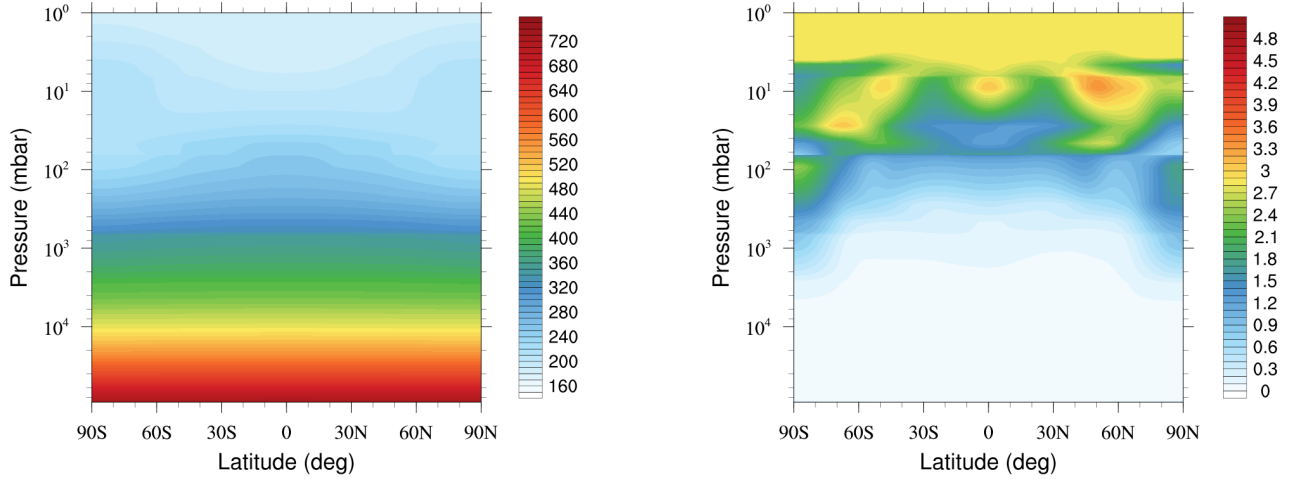

\begin{minipage}{0.48\linewidth}
\includegraphics[width=\linewidth]{plots/simulation-plots/temp_vertVslat_noAI.pdf}

\end{minipage}
\hfill
\begin{minipage}{0.48\linewidth}
\includegraphics[width=\linewidth]{plots/simulation-plots/dtemp_vertVslat.pdf}
\end{minipage}%
\label{fig:simulation-outputs}
\caption{\textbf{Left:} The above figure displays the simulated atmospheric temperature profile (in units of $\text{K}$) of Venus averaged over the final Venus solar day of the simulation. The simulation was run using OASIS and OASIS-RT for 5 Venus solar days (each Venus solar day is equivalent to approximately 117 Earth days). \textbf{Right:} The above figure displays the percentage difference in simulated atmospheric temperature profiles of Venus averaged across the final Venus day of the simulation, for two simulations using 1. OASIS-RT and 2. the surrogate models presented in this work, to model the radiative transfer. \textbf{Both:} In both plots, the main cloud deck extends from approximately $10^2 \, \text{mbar}$ to $2\times10^3 \, \text{mbar}$.}
\end{figure*}









\subsection{Model Performance on Test Set}
Table \ref{error-table} summarises the mean absolute error (MAE, equation \ref{eqn:mae-k}) across the four target variables, and relative to the mean values of these four variables, across the test set. These MAE values are in line with those achieved using similar surrogate modelling methods for radiative transfer within the Earth's atmosphere, with \cite{ukkonen_exploring_2022} quoting MAE for short-wave fluxes of around 1\% or less. For both the long-wave and short-wave regimes, the MAE is higher for the upwelling fluxes as compared to downwelling fluxes: this is expected as physical computation of the upwelling flux depends on the computation of the downwelling flux, thus meaning this is a more complicated mapping to emulate. The percentage errors for the short-wave targets are a factor of 2-3 greater than those for the long-wave targets: this is to be expected as the magnitude of the short-wave targets are smaller, and the supervised learning task set for the short-wave model in this work is a more complex mapping from inputs to outputs as compared to that for the long-wave model. 

In the following subsections, we visualise and interrogate the variation of model prediction errors with altitude and with scalar variables: $\cos\mu$ for both long-wave and short-wave model predictions, and surface temperature for long-wave model predictions only. For completeness, further plots of error as a function of the remaining input scalar variables (surface temperature, surface pressure, surface gas density) are contained in appendix section \ref{distribution-errors-scalar-parameters}.
\subsubsection{Short-Wave Surrogate Model}\label{short-wave-surrogate-model}
\indent

    A. \textit{Variation of error vs. altitude}
\vspace{0.2cm}
\newline
Figure \ref{fig:sw-mae} displays the average absolute error of predictions at different altitude levels across test samples. Below around $65\,\text{km}$, the average error increases roughly in proportion to the average target fluxes. Above this altitude, the average error decreases for both target variables, even though the average values of these variables continue to rise with altitude. This change in the trend of the average error occurs around the top of the cloud deck.

Potential explanations as to why there is lower error in predicting targets above $65\,\text{km}$ are as follows: above the top of the cloud deck, there is less complexity in mapping from input variables to output fluxes, and so this can be naively assumed to be an easier task to learn; there is also tighter variance in the target variables within the test set above this altitude threshold. 

Plots A and B of figure \ref{fig:sw-mae} display mean errors of less than 3\% across all altitude levels, averaged across test samples in the [5, 95] percentile interval of column-aggregated errors. This accuracy falls within an acceptably small margin of error. 
\vspace{0.2cm}
\newline
\indent
    B. \textit{Variation of error vs. $\cos\mu$}
\vspace{0.2cm}
\newline
Figure \ref{fig:error-variations} displays the distribution in test errors as a function of $\cos\mu$ of the test column. Data in these plots have been binned to more simply display the spread of errors for a given $\cos\mu$ interval. For values of $\cos\mu$ close to 1, there is a narrower distribution of error, with test samples being predicted reliably more accurately as compared to test samples corresponding to lower values of $\cos\mu$. 

This variation in test error distribution as a function of $\cos\mu$ may be attributable to the approximations used  during target preprocessing when training the model (see section \ref{data-preprocessing}); if indeed this is the case, then there is scope to mitigate against this by refining data preprocessing methods when producing future iterations of this surrogate model. This error variation with $\cos\mu$ may otherwise be due to the model not exactly capturing the complexities in how $\cos\mu$ is used in mapping inputs to outputs, which may be an acceptable and necessary trade-off in using a surrogate model for the purposes of model speed-up.

\subsubsection{Long-Wave Surrogate Model}
\vspace{0.2cm}

\indent
    A. \textit{Variation of error vs. altitude}
\vspace{0.2cm}
\newline
Figure \ref{fig:lw-mae} displays the absolute error of predictions averaged across altitude levels across test samples. A similar trend can be seen in these plots as compared to the trends described in \ref{short-wave-surrogate-model}: average magnitude of error roughly follows the same trend as average magnitude of the target variables, except for between $40-70 \, \text{km}$ altitude (roughly in the interval of the main cloud deck) whereby the error rises significantly at the top of the cloud deck and decreases going deeper into the cloud deck, for both target variables.

Below the cloud deck, it can be seen from plots A and B that the mean error in target predictions is less than 5\% of mean target magnitude, for the given percentile interval of test samples. Above the cloud deck, the long-wave fluxes tend to zero, so though the percentage errors increase for increasing altitude, these accuracies still fall within the reasonable margin of error for the model.
\vspace{0.2cm}
\newline
\indent
    B. \textit{Variation of error vs. $\cos\mu$/surface temperature}
\vspace{0.2cm}
\newline
Variations in test error versus surface temperature (displayed in figure \ref{fig:error-variations}), appear to follow the same trend as compared to residuals in the scaling factor used in preprocessing (displayed in figure \ref{fig:preprocessing-residuals}). This is promising as it may indicate that scaling residuals are the limiting factor in model accuracy, and these scaling residuals were initially deemed as acceptably small for the purpose of this work. Considering variation in test error versus cosine of the solar zenith angle ($\cos\mu$) (also displayed in figure \ref{fig:error-variations}), there is not a discernible trend in the error variation across $\cos\mu$, though it appears that the error distribution is most narrow for $\cos\mu$ approaching 1. This may reflect the variability in target flux profiles in both the train and test set for different $\cos\mu$ bins, as can be seen plotted in \ref{appendix-cosmu-variation}, whereby targets for higher values of $\cos\mu$ (i.e. approaching the substellar point) are more constrained, and thus are intuitively easier to predict.

\subsection{Model Performance in Simulation}\label{model-performance-in-simulation}

To evaluate the performance of our new surrogate model on 3D simulations of Venus's atmosphere, we have run two OASIS simulations: one using OASIS-RT for the radiative transfer scheme and the other using the new surrogate model presented in this work. In order to efficiently run the 3D simulation with OASIS-RT, as detailed in section \ref{oasis-rt}, we updated the radiative fluxes for solar radiation every 2880 steps, and the thermal radiation fluxes were updated every 320 steps. Both simulations run for 5 Venus solar days, each of which is approximately 117 Earth days. Figure \ref{fig:simulation-outputs} displays the temperature of the simulation using OASIS-RT averaged over the last simulated Venus day in the left-hand plot, and the percentage difference in this quantity between the two simulations in the right-hand plot. Beneath the bottom of the cloud deck, the percentage difference between the time-averaged temperature profiles produced by the two simulations is below 1.5\%; in the interval of the cloud deck, below 2.7\%; and above the cloud deck, below 4.0\%. These deviations in the final temperature profiles fall within the range of uncertainties in the measurements. This is reasonable due to two main factors: firstly, there will be inherent discrepancies between simulated and real temperatures arising from assumptions made in the physical model, and secondly, deviations will also naturally arise between the physical model and real temperature profiles due to the spatial resolution of the simulation. Also, we expected differences to arise due to the frequency at which the radiative fluxes are updated. In the case of the surrogate model, it is possible to update them at every physical time step. Section \ref{appendix-results} of the appendix displays the contrast of temperature profiles of simulations using OASIS-RT and the surrogate models, whereby the frequency of updating radiative fluxes in the latter matches that of the former (every 2880 steps for the short-wave regime and every 320 steps for the long-wave regime). There is negligible discrepancy between the final temperature profile of Venus as simulated using the surrogate-integrated schema for both frequencies of executing the radiative transfer update. A large difference between the plots was not expected because, in both cases, the radiative transfer updates are performed at intervals shorter than the radiative timescale. This means that similar results would occur if the simulation were executed using full radiative transfer updates at every timestep, as compared to the simulation executed using radiative transfer updates at every 2880/320 timesteps for SW/LW respectively. Despite this small difference in the resulting temperature profile corresponding to the different update frequencies, our future simulations will avoid using large step updates since these compromise the accuracy of the atmospheric physics, computed by the dynamical core, and would also compromise the future implementation of radiatively active clouds in the 3D simulations.






\subsection{Simulation Runtime}\label{sec:simulation-runtime}
To benchmark the speed-up achieved using the surrogate radiative transfer (RT) schema, we run four simulations of the Venus atmosphere. All four of these simulations span 1,000 timesteps, with each timestep corresponding to an increment of 15 seconds. Simulations were run on one NVIDIA Tesla V100 GPU. The simulations were as follows:

\begin{enumerate}
    \item Simulation with no RT update at any timestep: 95 seconds runtime
    \item Simulation with RT executed in the first timestep, and then updated every 2880 timesteps (equivalent to a timestep of 12 hours) for the short-wave regime, and every 320 timesteps (equivalent to a timestep of 1 hour 20 minutes) for the long-wave regime: 195 seconds
    \item Simulation with RT updated every timestep: 27,046 seconds
    \item Simulation using the surrogate models to update the RT every timestep: 184 seconds
\end{enumerate}

Simulation (ii) uses the typical RT update frequency employed for using OASIS to simulate Venus: this satisfies a trade off between sufficient temporal resolution of the RT, which must be modelled with a resolution smaller than the radiative time constant $\tau_R$ of the atmosphere ($\tau_R \approx 43 \text{ hours}$ at the altitude in the atmosphere where it takes its smallest value)\footnote{Radiative time constants for the Venus atmosphere can be seen in Table 2 of \cite{pollack_calculations_1975}} and overall simulation runtime. For clarity, in simulation (ii), the short-wave RT computation is executed once and the long-wave RT computation is executed 3 times. As mentioned in section \ref{model-performance-in-simulation}, less frequent RT updates limit the accuracy of modelling the atmospheric physics, and so a higher frequency of RT updates is preferred. Simulation (iii), which updates the RT at every 15 second timestep, illustrates how increasing the frequency of RT updates enormously increases the simulation runtime; here, by a factor of 92 compared to simulation (ii) with less frequent RT updates, and by a factor of 285 compared to simulation (i) with no RT update at all. 

Comparing simulation (iv) to simulation (ii) with infrequent RT updates, we see a speed-up of $\sim6\%$ whilst achieving a higher temporal resolution of RT. Comparing simulation (iv) to simulation (iii), we see a factor $147\times$ speed-up of the entire simulation runtime.

In addition to the reduction in simulation runtime, simulation (iv) is much more memory-efficient than simulations (i-ii), as the surrogate RT models do not require storing opacity cross-sections from the gas absorption or clouds. Furthermore, with the potential to update the radiative fluxes on every physical timestep now computationally feasible, our new approach allows for the inclusion of dynamical cloud feedback at a small computational cost. This addresses one of the main limitations of current 3D Venus atmospheric models.

\subsection{Limitations}

The surrogate models produced in this work do not take explicit input of the structure of cloud and gas absorber constituents of the atmosphere being modelled, except for the input of gas density, $\rho_{i,j}$. This means that the learning objective for surrogate models in this work is to approximate the mapping from input thermodynamic column variables to output columnar flux profiles, \textit{conditioned} on a specific cloud and absorber structure. Consequently, this means that the models produced in this work are only applicable for planets corresponding to the planetary parameters and cloud and absorber structure specific to Venus. This is a limitation in terms of \textit{generalisability} of the surrogate-integrated GCM to other types of atmosphere. 

\section{Conclusions}


This work introduces a surrogate model approach to replacing numerical simulations of short-wave and long-wave computations in a two-stream radiative transfer model, aimed at accelerating the Global Circulation Model (GCM), OASIS. The results show a significant GCM speed-up by a factor of 147 GPU performance, with surrogate models for both long-wave and short-wave regimes achieving test set accuracies of approximately 99\%. Additionally, this approach replicates the temperature profile of the original Venus simulations averaged across a Venus solar day with differences of 4\% after 5 Venus solar days of simulation.

This work is significant in that it enables:
\begin{enumerate}
\item $\sim150\times$ faster simulations of planets with massive atmospheres that require complex radiative transfer schemes, such as the Venus atmosphere.
\item Longer simulations with a much higher spatial resolution ($\sim10\times$).
\item Improved representation of the temperature evolution of short-term physical phenomena in the atmosphere. These can be atmospheric waves with timescales shorter than the period at which the radiative fluxes are updated in the simulation. In the case of our Venus simulations, we can measure the temperature change of atmospheric waves with timescales $<12$ hours.
\item A model free of model tuning to optimize performance, such as the frequency of how the radiative fluxes are updated.
\item The inclusion in 3D simulations of cloud dynamical feedback or higher-order, more complex radiative schemes with a small extra computational cost.
\end{enumerate}

This achievement of faster and/or higher spatial resolution atmospheric simulations will facilitate better insight into the nature of the atmosphere of Venus, as well as bench-marking the utility and applicability of such modelling techniques for use in exoplanet science.

\section*{Acknowledgements}\label{acknowledgements}
We thank Dr Ahmed Al-Refaie, Nikita Pond and Max Hart for helpful conversations on integrating TensorFlow models within a C/C++ framework.

The authors acknowledge the use of the High Performance Computing facilities of University College London to carry out this work, specifically the Hypatia cluster. This work used computing equipment funded by the Research Capital Investment Fund (RCIF) provided by UKRI, and partially funded by the UCL Cosmoparticle Initiative.
This research received funding from the European Research Council (ERC) under the European Union's Horizon 2020 research and innovation programme (grant agreement n$^\circ$ 758892/ExoAI), and from the Science and Technology Facilities Council (STFC; grant n$^\circ$ ST/W00254X/1 and grant n$^\circ$ ST/W50788X/1). JMM acknowledges support from the Horizon Europe Guarantee Fund, grant EP/Z00330X/1.

\section*{Data Availability}\label{code}


The code for this project will be made publicly available at https://github.com/ttahseen/oasis-rt-surrogate on acceptance of this paper.



\bibliographystyle{mnras}

\bibliography{references} 




\onecolumn
\appendix
\section{Model Architecture}\label{appendix:model-architecture}
\subsection{Feedforward Neural Networks}

A \textit{neural network} is a type of function $f$ which, given an input feature vector $\mathcal{X}$, can be implemented as combination of matrix transformations, $\{\mathcal{M}_k \in \mathbb{R}^{m_k \times n_k}\}$, and non-linear transformations, $\{\alpha_k\}$, where each $\alpha_k$ is a dimension-preserving mapping, i.e. $\alpha_k: \mathbb{R}^{m \times n} \rightarrow \mathbb{R}^{m \times n}$. A simple feedforward neural network implements the transformation\footnote{Below, \( \bigcirc_{k=0}^{i} (a_k \circ M_k) \) represents the composition of functions \( (a_k \circ M_k) \) from \( k = 0 \) to \( k = i \).}
\begin{equation}
    \mathcal{Y} = \left( \bigcirc_{k=0}^{i} (\alpha_k \circ \mathcal{M}_k) \right)(\mathcal{X})
\end{equation}
The matrices $\{\mathcal{M}_k\}$ are fitted as an optimisation task of a chosen objective function $\mathcal{L}$ given a set of examples $\mathcal{S} = \{\hat{\mathcal{Y}}, \hat{\mathcal{X}}\}$ to constrain the network to approximate some given mapping $\tilde{f}$ captured by $\mathcal{S}$. Neural networks as described above are universal function approximators, and so constitute an appropriate function space within which to seek an approximate function $f$ for the target function $\tilde{f}$. \\The matrices $\{\mathcal{M}_k\}$ are often referred to as network \textit{weights} or network \textit{parameters}; the dimension of each $\mathcal{M}_k$ is free in one dimension (except for $\mathcal{M}_i$ for which the dimension is fully constrained by the dimension of $\mathcal{Y}$ and the dimension of $\left( \bigcirc_{k=0}^{i-1} (\alpha_k \circ \mathcal{M}_k) \right)(\mathcal{X})$). This free dimension $m_k$ of each  $\mathcal{M}_k$ we can ascribe as the number of \textit{neurons} of the \textit{layer} $\mathcal{M}_k$, to be consistent with machine learning terminology. For further details about the mathematics and implementation of neural networks, see chapters 3 \& 4 of \cite{prince_understanding_2023}.


\subsection{Recurrent Neural Networks}

Recurrent neural networks (RNNs) are function spaces defined using similar concepts to simple feedforward neural networks, designed to incorporate the structural dependence of sequential input feature vectors. Given a set of input feature vectors $\{X_k\} \, \forall \, k \in [0, N]$, with some relation between consecutive feature vectors $X_k$ and $X_{k+1}$, a simple recurrent neural network operates as follows:

\begin{equation}
    r_{k+1} = \alpha_0(\mathcal{M}_{\text{in}} \, \mathcal{X}_{k+1} + \mathcal{M} r_k)
\end{equation}
\begin{equation}
    \mathcal{Y}_{k+1} = \alpha_1 (\mathcal{M}_{\text{out}} \, r_{k+1})
\end{equation}
\noindent where $r_k$ is defined as the \textit{hidden state} of the network for the $k$th input. The above formulation of the simple recurrent network illustrates how information from preceding input feature vectors contributes to the function output corresponding to the $k$th input.

In practice, the RNN implementation is more complicated than the two equations above, but these equations capture the core operation of the recurrent neural network. For an in-depth treatment of the mathematics and implementation pragmatics of recurrent neural networks, see \cite{geron_chapter_2019}.

\newpage

\section{Results}\label{appendix-results}

Below we display the contrast of temperature profiles of simulations using OASIS-RT and the surrogate models, whereby the frequency of updating radiative fluxes in the latter matches that of the former (every 2880 steps for the short-wave regime and every 320 steps for the long-wave regime). There is negligible discrepancy between the final temperature profile of Venus as simulated using the surrogate-integrated schema for both frequencies of executing the radiative transfer update.

\begin{figure}[H]
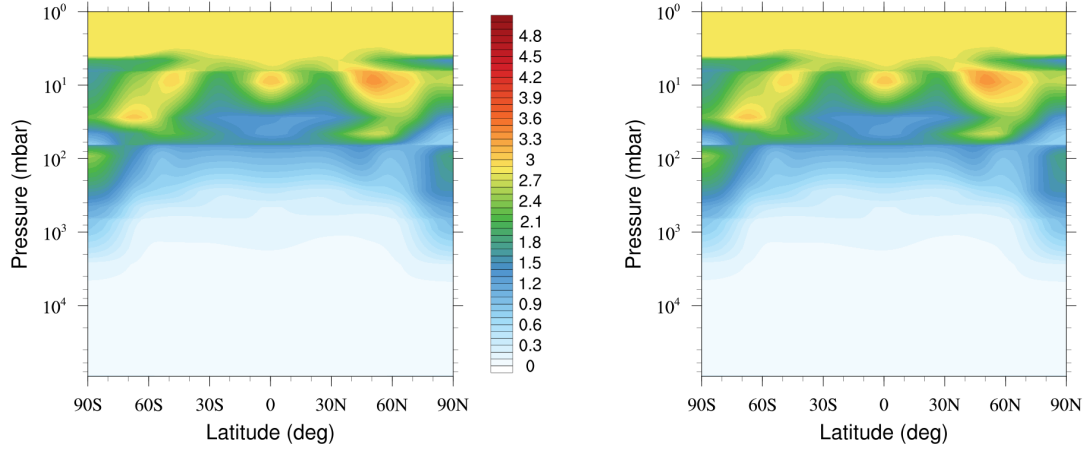

\centering
\includegraphics[width=0.45\textwidth]{plots/simulation-plots/dtemp_vertVslat_v2.pdf}
\includegraphics[width=0.45\textwidth]{plots/simulation-plots/dtemp_vertVslat.pdf}
\caption{\textbf{Both:} The above figures display the percentage difference in simulated atmospheric temperature profiles of Venus averaged across the final Venus day of the simulation, for two simulations using 1. OASIS-RT and 2. the surrogate models presented in this work, to model the radiative transfer. \textbf{Left: } This plot was generated by updating the short-wave radiative fluxes at every 2880 steps, and the long-wave radiative fluxes at every 320 steps, for the simulation with the surrogate models. \textbf{Right: } This plot was generated by updating the short-wave radiative fluxes and long-wave radiative fluxes at every timestep, for the simulation with the surrogate models.}
\label{fig:simulation-results-v2}
\end{figure}

\newpage

\section{Model Parameters}\label{appendix-model-parameters}

The tables below display the number of model parameters per layer of the two surrogate models produced in this work.

\subsection{Short-Wave Surrogate Model}

\begin{table}[H]
\begin{center}
    \begin{tabular}{|c|c|c|} \hline  
         \textbf{Layer}&  \textbf{Output Shape}& \textbf{Number of Parameters}\\ \hline  
         Main Inputs&  [($n$, 49, 3)]& 0\\ \hline  
         Auxiliary Inputs&  [($n$, 5)]& 0\\ \hline  
 $\text{GRU}_{\downarrow}$& [($n$, 49, 128), ($n$, 128)]&51,072\\ \hline 
 $\text{Dense}_{1}$& [($n$, 128)]&17,152\\ \hline  
 $\text{GRU}_{\uparrow}$& [($n$, 50, 128)]&99,072\\ \hline 
 $\text{Dense}_{\text{out}}$& [($n$, 50, 2)]&514\\ \hline 
 \multicolumn{2}{|c|}{\textbf{Total number of model parameters:}}&167,810\\ \hline 
    \end{tabular}
    \end{center}
    \caption{This table displays the number of parameters across model layers for the short-wave surrogate model, as well as the output shapes of each layer. $n$ denotes the number of atmospheric columns passed as input to the model; layers correspond to those displayed in figure \ref{fig:model-architecture}.}
    \label{tab:sw-surrogate-model-parameters}
\end{table}

\subsection{Long-Wave Surrogate Model}
\begin{table}[H]
\begin{center}
    \begin{tabular}{|c|c|c|} \hline  
         \textbf{Layer}&  \textbf{Output Shape}& \textbf{Number of Parameters}\\ \hline  
         Main Inputs&  [($n$, 49, 3)]& 0\\ \hline  
         Auxiliary Inputs&  [($n$, 4)]& 0\\ \hline  
 $\text{GRU}_{\downarrow}$& [($n$, 49, 32), ($n$, 32)]&3,552\\ \hline 
 $\text{Dense}_{1}$& [($n$, 32)]&1,184\\ \hline  
 $\text{GRU}_{\uparrow}$& [($n$, 50, 32)]&6,336\\ \hline 
 $\text{Dense}_{\text{out}}$& [($n$, 50, 2)]&130\\ \hline 
 \multicolumn{2}{|c|}{\textbf{Total number of model parameters:}}&11,202\\ \hline 
    \end{tabular}
    \end{center}
    \caption{This table displays the number of parameters across model layers for the long-wave surrogate model, as well as the output shapes of each layer. $n$ denotes the number of atmospheric columns passed as input to the model; layers correspond to those displayed in figure \ref{fig:model-architecture}.}
    \label{tab:lw-surrogate-model-parameters}
\end{table}

\newpage
\section{Further aggregate statistics of model test set performance}

\begin{figure}[H]
\includegraphics[width=1\linewidth]{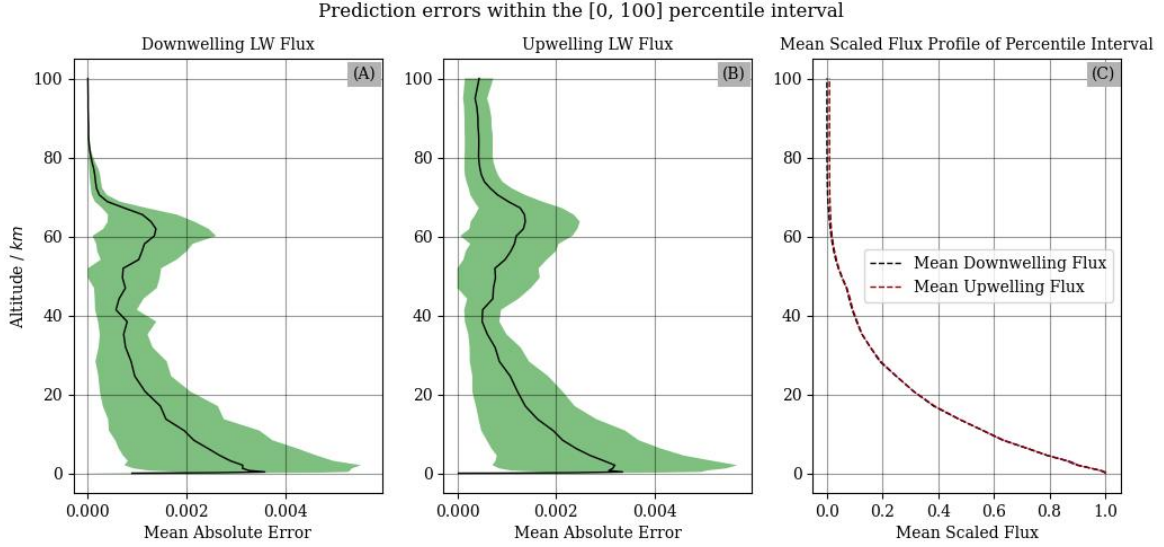}
\caption{The above plots display the $\widehat{\text{MAE}}_{j,k}$ (defined in equation \ref{scaled_MAE_jk}) of the long-wave surrogate model predictions. Plot C displays the target scaled flux profiles averaged across the test set. Plot F displays the target flux profiles averaged across the test set. Plots A and B display the mean absolute error of the scaled predictions and targets across the test set. Plots D and E display the mean absolute error of the unscaled predictions and targets across the test set.}
\label{fig:lw-mae-0-100}
\end{figure}

\begin{figure}[H]
\includegraphics[width=1\linewidth]{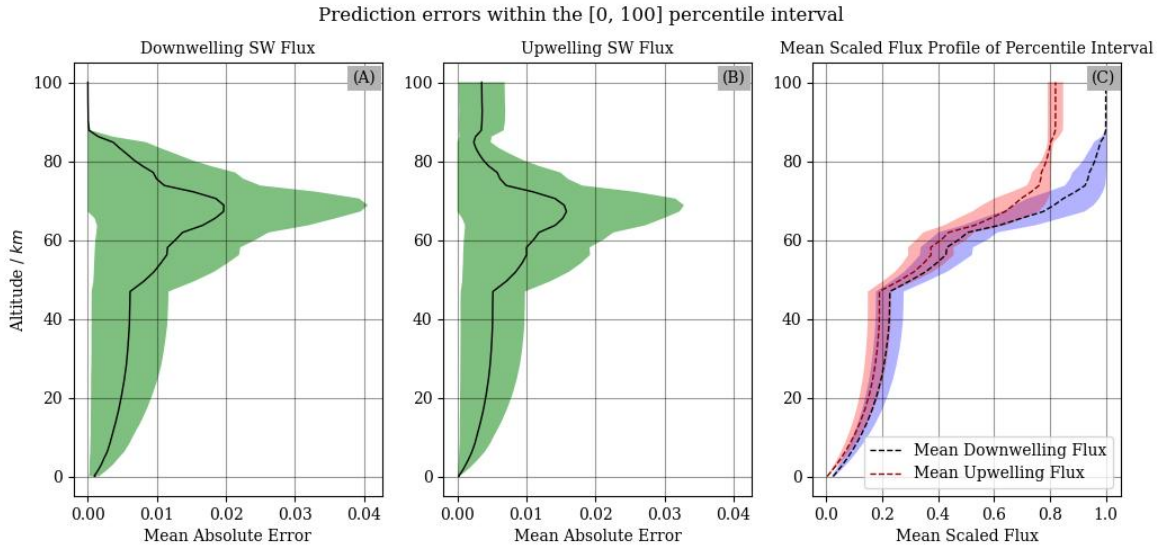}
\caption{The above plots display the $\widehat{\text{MAE}}_{j,k}$ (defined in equation \ref{scaled_MAE_jk}) of the short-wave surrogate model predictions. Plot C displays the target scaled flux profiles averaged across the test set. Plot F displays the target flux profiles averaged across the test set. Plots A and B display the mean absolute error of the scaled predictions and targets across the test set. Plots D and E display the mean absolute error of the unscaled predictions and targets across the test set.}
\label{fig:sw-mae-0-100}
\end{figure}

The figures included in this section display the $\widehat{\text{MAE}}_{j,k}$ (defined in equation \ref{scaled_MAE_jk}) the entire test set for the long-wave (Figure \ref{fig:lw-mae-0-100}) and short-wave (Figure 
\ref{fig:sw-mae-0-100}) schemas. These figures are included for completeness, and are analogous to plots A-C of \ref{fig:sw-mae} and \ref{fig:lw-mae} but for the entire test sets instead of the [5, 95] percentile interval subset of the test sets.

\newpage
\section{Distribution of Samples}\label{distribution-of-samples}
The plots in this section display the distribution of the entire data set across surface temperature and cosine of the solar zenith angle.

\begin{figure}[h]
\centering
\begin{minipage}{0.42\linewidth}
\includegraphics[width=\linewidth]{plots/new-paper-plots/01-jul-2024/surface-temperature-distribution.pdf}
\end{minipage}
\hfill
\begin{minipage}{0.42\linewidth}
\includegraphics[width=\linewidth]{plots/new-paper-plots/01-jul-2024/cosz-distribution.pdf}
\end{minipage}
\begin{minipage}{0.42\linewidth}
\includegraphics[width=\linewidth]{plots/new-paper-plots/01-jul-2024/stemp-dayside-distribution.pdf}
\end{minipage}
\hfill
\begin{minipage}{0.42\linewidth}
\includegraphics[width=\linewidth]{plots/new-paper-plots/01-jul-2024/stemp-nightside-distribution.pdf}
\end{minipage}%
\label{fig:sample-distributions}
\caption{
\textbf{Top row:} The figure displays the distributions of surface temperature (left tile) and cosine of the solar zenith angle, $\cos\mu$ (right tile) across the dataset used for this work.
\\ 
\textbf{Bottom row:} The figure displays the distributions of surface temperature across samples falling within the dayside (left tile) and nightside (right tile).
}
\end{figure}

\begin{figure}[H]
\centering
\begin{minipage}{0.5\linewidth}
\includegraphics[width=\linewidth]{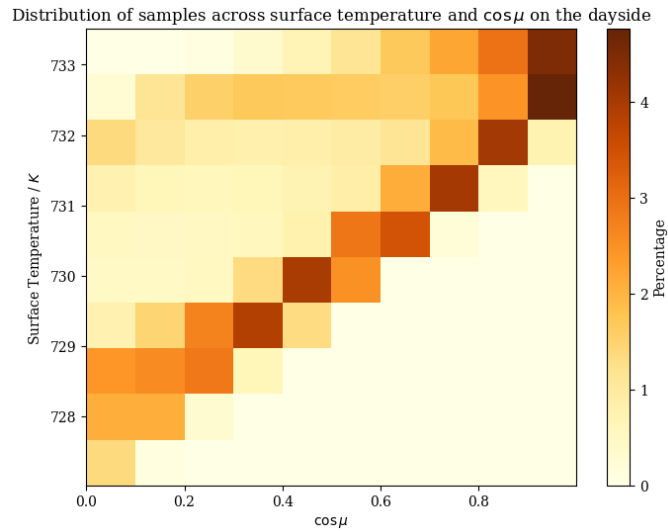}
\end{minipage}
\caption{The figure displays the co-variation of surface temperature and $\cos\mu$ across dayside samples within the dataset.}
\end{figure}

\newpage
\section{Distribution of Errors with Scalar Parameters}\label{distribution-errors-scalar-parameters}

\begin{figure}[H]
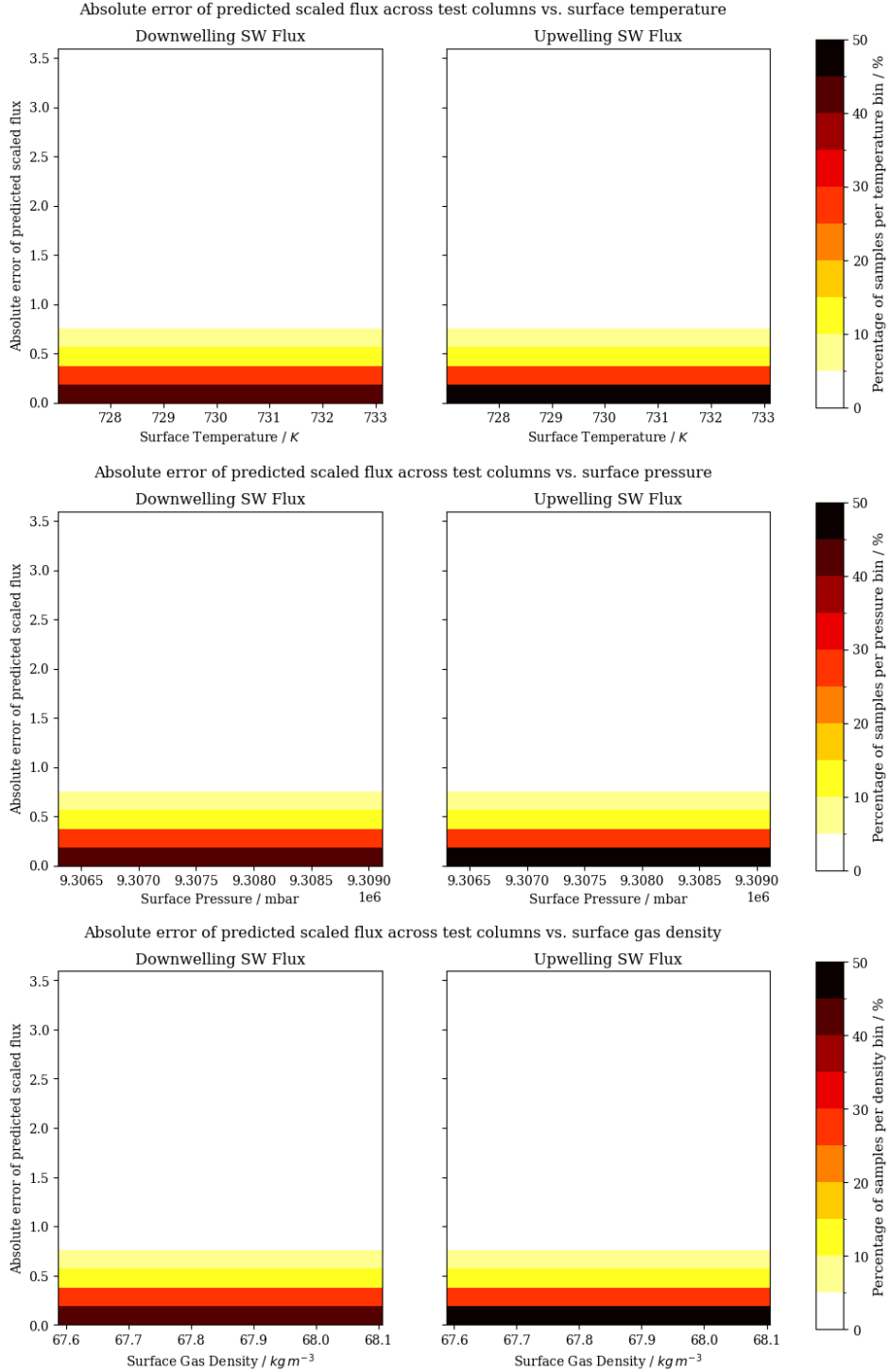

\centering
\includegraphics[width=0.7\textwidth]{plots/new-paper-plots/01-jul-2024/sw-err-stemperature.pdf}
\includegraphics[width=0.7\textwidth]{plots/new-paper-plots/01-jul-2024/sw-err-spressure.pdf}
\includegraphics[width=0.7\textwidth]{plots/new-paper-plots/01-jul-2024/sw-err-srho.pdf}
\caption{The figure displays variation in $\widehat{\text{CAE}}_{i,k}$ (as defined in equation \ref{scaled_CAE_ik}) of short-wave model predictions as a function of surface temperature (top row), surface pressure (middle row), and surface gas density (bottom row). The extent of the plots on the y-axis covers the full range of the error quantity on the y-axis of each plot, with white spaces corresponding to values falling in the lowest value bin. Bins have been normalised to percentages by the total number of test columns in each x-axis bin.}
\label{fig:sw-error-variations}
\end{figure}

\begin{figure}[H]
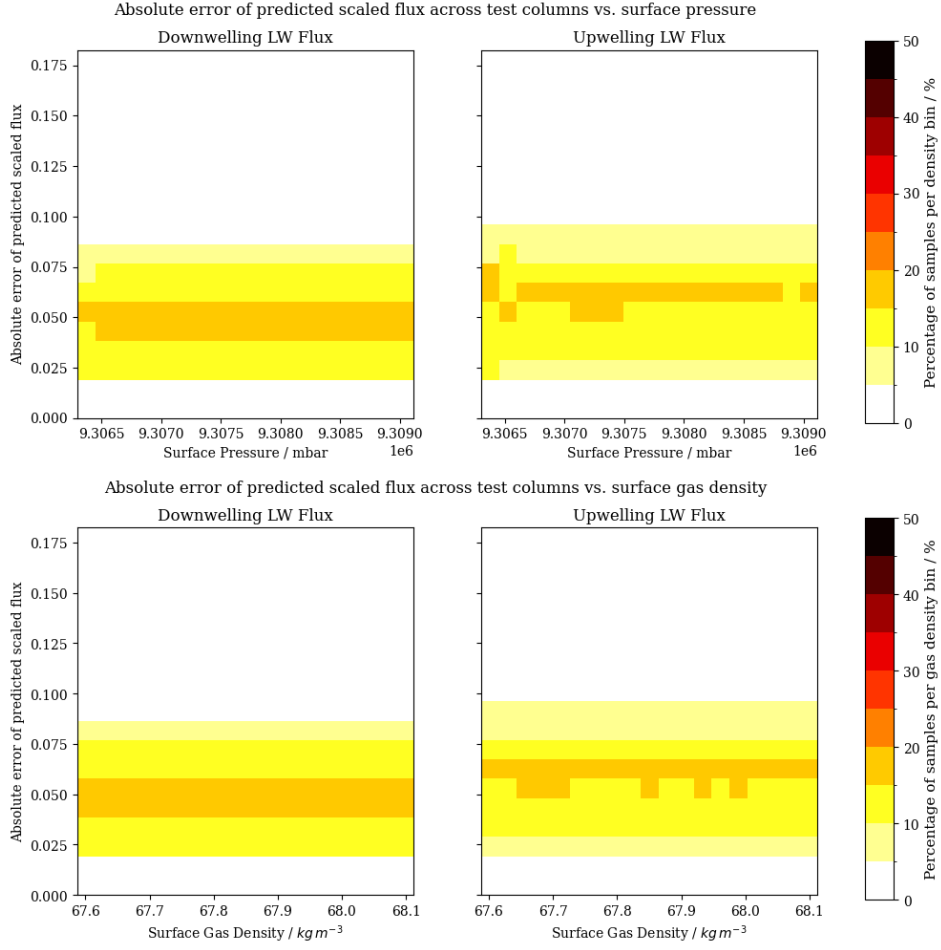

\begin{center}
\includegraphics[width=0.7\textwidth]{plots/new-paper-plots/01-jul-2024/lw-err-spressure.pdf}
\includegraphics[width=0.7\textwidth]{plots/new-paper-plots/01-jul-2024/lw-err-srho.pdf}
\end{center}
\caption{The figure displays variation in $\widehat{\text{CAE}}_{i,k}$ (as defined in equation \ref{scaled_CAE_ik}) of long-wave model predictions as a function of surface pressure (top row), and surface gas density (bottom row). The extent of the plots on the y-axis covers the full range of the error quantity on the y-axis of each plot, with white spaces corresponding to values falling in the lowest value bin. Bins have been normalised to percentages by the total number of test columns in each x-axis bin.}
\label{fig:lw-error-variations}
\end{figure}

The plots in this section are analogous to those displayed in Figure \ref{fig:error-variations} and display the variation in $\widehat{\text{CAE}}_{i,k}$ (as defined in equation \ref{scaled_CAE_ik}) of short-wave (Figure \ref{fig:sw-error-variations}) and long-wave model predictions (Figure \ref{fig:lw-error-variations}) versus scalar model variables. These figures are included for completeness.

\newpage
\section{Variation of error vs. cos$\mu$ in the short-wave regime}\label{appendix-cosmu-variation}
\begin{figure}[H]
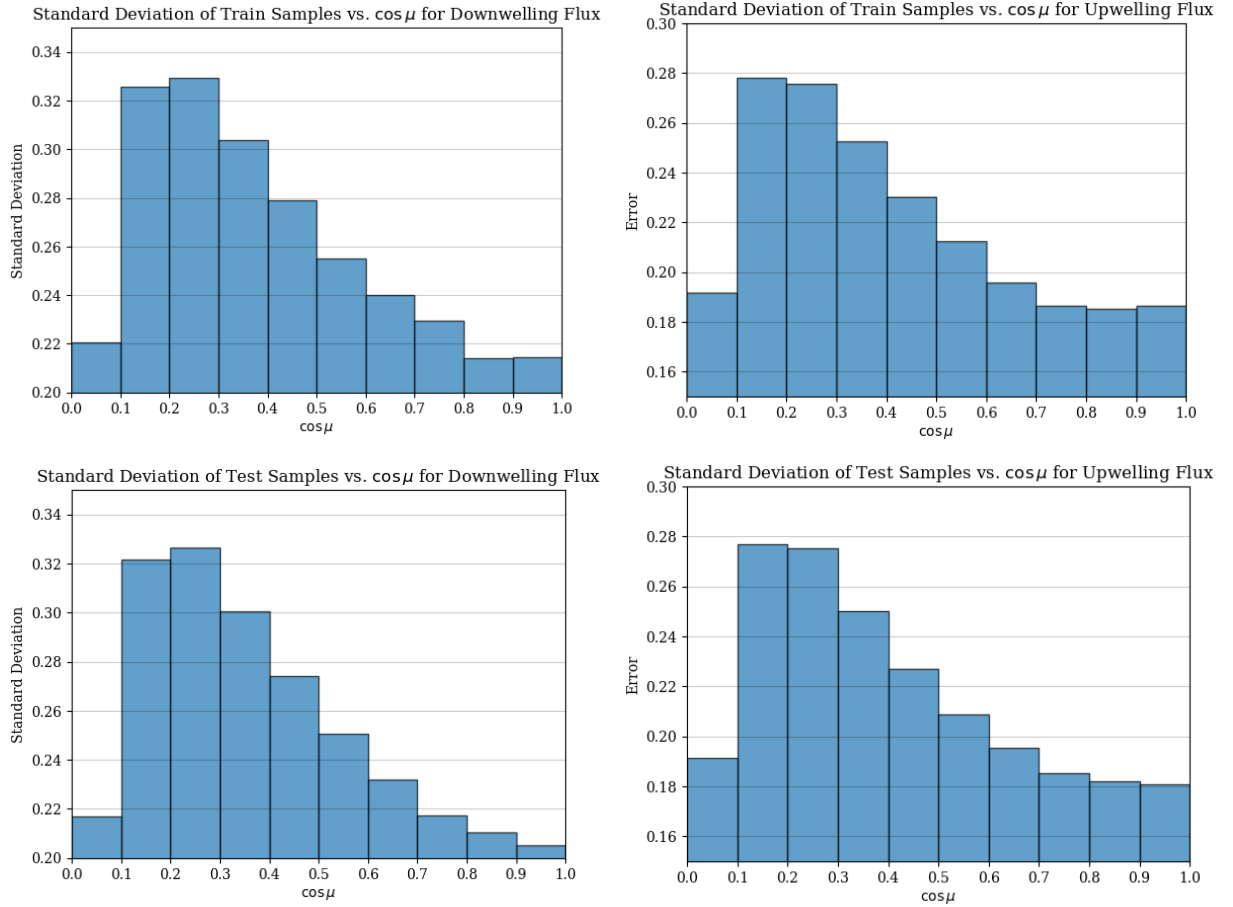

\centering
\begin{minipage}{0.45\linewidth}
\includegraphics[width=\linewidth]{plots/appendix-error-variation-cos-train-downwelling.pdf}
\end{minipage}
\begin{minipage}{0.45\linewidth}
\includegraphics[width=\linewidth]{plots/appendix-error-variation-cos-train-upwelling.pdf}
\end{minipage}%
\hfill
\begin{minipage}{0.45\linewidth}
\includegraphics[width=\linewidth]{plots/appendix-error-variation-cos-test-downwelling.pdf}
\end{minipage}
\begin{minipage}{0.45\linewidth}
\includegraphics[width=\linewidth]{plots/appendix-error-variation-cos-test-upwelling.pdf}
\end{minipage}
\caption{The above figure shows the distribution of column-aggregated error $\sigma_{i, k}$ (computed according to equation \ref{sigma_column}) per $\cos\mu$ bin per short-wave target for both the train set (top row) and test set (bottom row). These figures illustrate a lower level of variation in the flux-altitude profiles with higher values of $\cos\mu$, and for the bin with the lowest values of $\cos\mu$ (i.e. at the terminator).}
\label{fig:f1}
\end{figure}

To investigate the amount of variability in targets versus $\cos\mu$, we define the following quantity:
\begin{equation}
    \sigma_{i, k} = \sum_{j=0}^{n_{\text{levels}}-1} \sigma_{i,j,k}
    \label{sigma_column}
\end{equation} and plot this quantity for the test set, displayed in Figure \ref{fig:f1}. The plots displayed in Figure \ref{fig:f1} illustrate a lower level of variation in the flux-altitude profiles with higher values of $\cos\mu$, and for the bin with the lowest values of $\cos\mu$ (i.e. at the terminator).

\bsp	
\label{lastpage}
\end{document}